\documentclass[twocolumn,preprintnumbers,prx]{revtex4}
\usepackage{graphicx}
\usepackage{dcolumn}
\usepackage{amssymb}
\usepackage{bm}

\begin{document}

\title{Exactly solvable model of Fermi arcs and pseudogap}
\author{Kun Yang}
\affiliation{Department of Physics and National High Magnetic Field Laboratory,
Florida State University, Tallahassee, Florida 32306, USA}
%\pacs{73.43.Nq, 73.43.-f}

\begin{abstract}

We introduce a very simple and exactly solvable model that supports Fermi arcs in its ground state and excitation spectrum. These arcs come in pairs, and merge into what we call a pseudo Fermi surface along which fermions are gapped; this fermion gap is naturally identified as a pseudogap. Comparison will be made with phenomenology of high temperature cuprate superconductors.

\end{abstract}

\date{\today}

\maketitle

\section{Introduction}

High transition temperature ($T_c$) cuprate superconductors remain a central topic of condensed matter research since their discovery in the 1980's.
Among their many fascinating and mysterious properties\cite{Proust}, the existence of pseudogaps and Fermi arcs in the underdoped regions of their phase diagrams is particularly intriguing\cite{Yoshida}. Both of them indicate the breakdown of the standard Fermi liquid\cite{Book} description of the non-superconducting state above $T_c$, on which the Bardeen-Cooper-Schrieffer (BCS) theory\cite{Book} of the superconducting state is founded. Such non-Fermi liquid behavior is also seen in many other properties including lack of coherence in quasiparticle excitations and unusual transport properties in the normal state\cite{Proust,Yoshida}, which are likely related to the physics of neighboring Mott insulator phase. Undertanding such non-Fermi liquid physics, including (and in particular) the origin of Fermi arcs and pseudogap, is an exciting challenge we face.

In this paper we present an extremely simple and exactly solvable model, and show that Fermi arcs and pseudogap appear very naturally (and hand-in-hand) in its ground state and excitation spectrum. Our model is a variant of a model introduced by Hatsugai and Kohmoto (HK)\cite{hk} (a model similar to that of HK was considered earlier by Baskaran\cite{Baskaran}).
%The HK model leads to an exact solution for the Mott-insulating ground state at half-filling, which corresponds to the undoped parent compound the cuprates. Thus combining our results with those of HK's offers a unified description of the Mott-insulting, pseudogap and Fermi arc phenomena.
We note the HK model is also the basis of a recent theory of superconductivity in doped Mott insulators\cite{Phillips}, building on earlier phenomenological work\cite{Setty}. We would like to emphasize upfront that it is {\em not} our purpose to present a comprehensive theory for the cuprate phenomenology based on such a simple model, which, as we will explain in more detail below, misses certain important aspects of the cuprate physics. Its value is providing a proof of principle, namely Fermi arcs and pseudogap can exist. Given their importance, we believe a simple model that provides an existence proof is highly desirable. It may also provide a starting point for building more realistic models that can eventually lead to a comprehensive understanding of the extremely rich phenomenology of the cuprates.

\section{Model and its solution}

We consider the following model:
\begin{equation}
H=\sum_{{\bf k}}[\epsilon_{{\bf k}}(n_{{\bf k}\uparrow} + n_{{\bf k}\downarrow})+u_{{\bf k}} n_{{\bf k}\uparrow}n_{{\bf k}\downarrow}],
\label{eq:model}
\end{equation}
where the single particle energy $\epsilon_{{\bf k}}$ is measured from the chemical potential, and $n_{{\bf k}\sigma}$ is the fermion occupation for momentum ${\bf k}$ and spin $\sigma = \uparrow,\downarrow$. For simplicity and concreteness, we consider, {\em e.g.}, a square lattice with (anisotropic) nearest neighbor hopping, resulting in
\begin{equation}
\epsilon_{{\bf k}} =-2t_x\cos(k_x) - 2t_y\cos(k_y) -\mu,
\label{eq:dispersion}
\end{equation}
where $t_x$ and $t_y$ are hopping along x- and y-directions, and $\mu$ is chemical potential.
If the interaction energy $u_{{\bf k}} = U$ is a constant, (\ref{eq:model}) reduces to the HK model\cite{hk}. The momentum dependence of $u_{{\bf k}}$ breaks the very special center of mass conservation of the HK interaction, and can be understood as renormalization of dispersion of the interacting pair. Since such interaction is expected to emerge at low energy from the interplay of single particle dispersion and more realistic two-body interaction, existence of such dispersion renormalization is physically very natural. Again for simplicity and concreteness, we consider
\begin{equation}
u_{{\bf k}} =-2T_x\cos(k_x) - 2T_y\cos(k_y) +U,
\label{eq:pairdispersion}
\end{equation}
where $T_x$ and $T_y$ may be understood as the renormalization of hopping of center of mass of the interacting pair.
The solvability of the model lies in the fact that $n_{{\bf k}\uparrow}$ and $n_{{\bf k}\downarrow}$ are conserved quantities. We emphasize the details of the momentum dependence of Eqs. (\ref{eq:dispersion}) and (\ref{eq:pairdispersion}) are {\em not} important for our discussions below.

If $u_{{\bf k}}$ is either positive or negative definite, the solution is qualitatively very similar to that of the HK model. In the following we show the situation is very different when $u_{{\bf k}}$ changes sign in momentum space. In particular when the $u_{{\bf k}}=0$ surface intersects with the non-interacting Fermi surface $\epsilon_{{\bf k}} = 0$, the ground state supports Fermi arcs.

Defining $n_{{\bf k}} = n_{{\bf k}\uparrow} + n_{{\bf k}\downarrow}$, we find in the ground state
\begin{eqnarray}
n_{{\bf k}}= \left\lbrace\begin{array}{cc}
0, &\epsilon_{{\bf k}} > 0 \hskip 0.2 cm {\rm and} \hskip 0.2 cm 2\epsilon_{{\bf k}} + u_{{\bf k}} > 0;\\
1, &\epsilon_{{\bf k}} < 0 \hskip 0.2 cm {\rm and} \hskip 0.2 cm \epsilon_{{\bf k}} + u_{{\bf k}} > 0;\\
2, &\epsilon_{{\bf k}} + u_{{\bf k}} < 0 \hskip 0.2 cm {\rm and} \hskip 0.2 cm 2\epsilon_{{\bf k}} + u_{{\bf k}} < 0. \end{array}\right.
\label{eq:occupation}
\end{eqnarray}
The surfaces (or lines) in momentum space across which $n$ jumps are given by
\begin{eqnarray}
\label{eq:boundary 1}
&&\epsilon_{{\bf k}} = 0,\\
\label{eq:boundary 2}
&&\epsilon_{{\bf k}} + u_{{\bf k}} = 0,\\
\label{eq:boundary 3}
&& 2\epsilon_{{\bf k}} + u_{{\bf k}} = 0.
\label{eq:boundaries}
\end{eqnarray}
Obviously all three lines intersect with the loop
\begin{eqnarray}
u_{{\bf k}}=0
\label{eq:boundary 4}
\end{eqnarray}
at the same points. In Fig. 1(a) all four lines are plotted in a simple representative case, resulting in the occupation pattern Fig. 1(b).

\begin{figure}%[hbt]
%\vspace{2in}
\center
\includegraphics[width=0.45\textwidth]{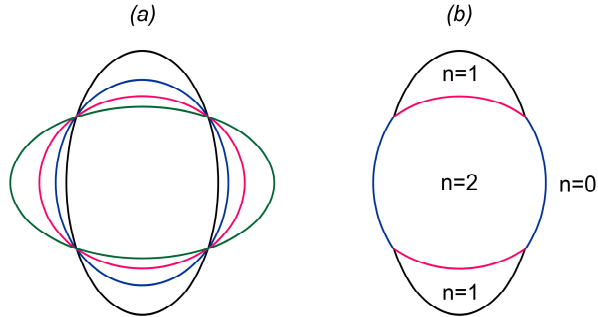}
\caption{(a) Solutions of Eqs. (\ref{eq:boundary 1}) (black line), (\ref{eq:boundary 2}) (red line), (\ref{eq:boundary 3}) (blue line) and (\ref{eq:boundary 4}) (green line). We assume the quantities that vanish along these loops are negative inside and positive outside. (b) Occupation pattern corresponding to panel (a). The black and red lines are Fermi arcs while the blue lines are pseudo Fermi surfaces along which there is a pseudogap.}
\label{fig:fig1}
\end{figure}

We now distinguish two different types of boundaries in Fig. 1(b).

$\bullet$ $\Delta n_{{\bf k}} =2$. This case looks (superficially) like the Fermi surface of the non-interacting Fermi gas, but is {\em not} a Fermi surface, because single fermion excitations are actually {\em gapped} there with gap $|u_{{\bf k}}|/2$.
%Instead singlet fermion {\em pairs} are gapless there;
As a result we refer to such boundaries pseudo Fermi surfaces, and identify the fermion gap there the pseudogap (to be distinguished from superconducting gap; more on this point later).

$\bullet$ $\Delta n_{{\bf k}} =1$. This case corresponds to our Fermi arcs, and are Fermi surfaces in the sense single fermion excitations are gapless there.

%{\em Discussions} ---
Having identified the Fermi arcs and pseudogap regions in our model, a few comments are in order.

$\bullet$
As pointed out by HK\cite{hk} the region with $n_{{\bf k}}=1$ (surround by the Fermi arcs in our case) has a massive degeneracy associated with a two-fold degeneracy for each ${\bf k}$. This is an artifact of the model. In fact this degeneracy can be easily removed by introducing a Zeeman splitting $\Delta_Z$ between up- and down-spin fermions:
\begin{eqnarray}
H_Z&=&\Delta_Z\sum_{{\bf k}}n_{{\bf k}\uparrow}\\
&=&\frac{\Delta_Z}{2}\sum_{{\bf k}}(n_{{\bf k}\uparrow} - n_{{\bf k}\downarrow})+\frac{\Delta_Z}{2}\sum_{{\bf k}}(n_{{\bf k}\uparrow} + n_{{\bf k}\downarrow}),
\label{eq:Zeeman}
\end{eqnarray}
and our previous analysis goes through exactly the same way, as long as one replaces $u_{{\bf k}}$ by $u_{{\bf k}} + \Delta_Z$ (note the last term above results in a shift of chemical potential $\mu$). Then the two Fermi arcs surrounding the $n_{{\bf k}}=1$ region are Fermi surfaces for the up- and down-spin fermions respectively. Even if $\Delta_Z=0$, as we will demonstrated later, ferromagnetic (Stoner) instability will remove this degeneracy when generic weak repulsive interaction is present. For the ease of discussion we will assume such a splitting is present (regardless of its origin), although most of our conclusions hold even when the massive degeneracy mentioned above is present.

$\bullet$ The pseudogap is clearly a singlet pairing gap. However it is {\em not} due to a fluctuating BCS pairing order parameter. In fact to (literally) the opposite, the pairing here is between up- and down-spin fermions at the {\em same} momentum, instead of opposite momenta as in BCS. The pairing here could perhaps be viewed as an extreme version of fluctuating pair density wave (PDW)\cite{Agterberg}, in which {\em every} pair has a different (and very large) momentum. We note a fluctuating PDW has been proposed by Lee and co-workers\cite{Lee} to be the mother of many competing states in underdoped cuprates, and hold the key to Fermi arcs and pseudogap.

$\bullet$ As discussed earlier, Zeeman splitting results in an effective increase of $u_{{\bf k}}$. This widens the Fermi arcs at the expense of the pseudo Fermi surface (or pseudogap region). More quantitatively, the pseudogap as well as pseudo Fermi surface are eliminated when
\begin{eqnarray}
\Delta_Z = |u_{\rm min}|=\Delta_{\rm pseudo},
\label{eq:Zeeman}
\end{eqnarray}
where $u_{\rm min}$ is the (negative) minimum value of $u_{{\bf k}}$ along the pseudo Fermi surface, whose magnitude is the maximum value of pseudogap $\Delta_{\rm pseudo}$.
Such suppression of pseudogap by Zeeman splitting, and in particular the simple linear scaling with $\Delta_Z$, have indeed been seen in the cuprates\cite{Shibauchi}.

\section{Angle-resolved photoemission spectroscopy and Luttinger sum rule}

An angle-resolved photoemission spectroscopy (ARPES) experiment\cite{Yoshida} measures the negative energy portion ($\omega < 0$, representing states occupied by electrons) of the electron spectral function, which is the imaginary part of the retarded electron Green's function:
\begin{eqnarray}
G(\omega, {\bf k}) = \frac{1}{\omega - \epsilon_{\bf k} + i\delta}
%\label{eq:Zeeman}
\end{eqnarray}
if $n_{\bf k} = 0$ in the ground state,
\begin{eqnarray}
G(\omega, {\bf k}) = \frac{1}{\omega - \epsilon_{\bf k} - u_{\bf k} + i\delta}
%\label{eq:Zeeman}
\end{eqnarray}
if $n_{\bf k} = 2$ in the ground state, and
\begin{eqnarray}
G_\uparrow(\omega, {\bf k}) &=& \frac{1}{\omega - \epsilon_{\bf k} - u_{\bf k} + i\delta},\\
G_\downarrow(\omega, {\bf k}) &=& \frac{1}{\omega - \epsilon_{\bf k} + i\delta}
\end{eqnarray}
if $n_{\bf k} = 1$ in the ground state. In the above $\delta$ is an infinitesimal positive, and we took the limit $\Delta_Z\rightarrow 0$ so that we do not need to distinguish between up- and down-spin Green's function in regions with $n_{\bf k} = 0$ and $n_{\bf k} = 2$ in the ground state. None of the conclusions below depends on this simplification. Due to the single pole structure, the spectral function contains a single delta-function peak in our model.

ARPES will find the Fermi arcs exactly the same way ordinary Fermi surfaces are identified. It will (in principle) see both branches (and therefore a closed surface) if it is not spin-resolved, although not necessarily resolving them if their separation is smaller than the momentum resolution of ARPES. On the other hand a spin-resolved ARPES will see one branch but not the other, resulting in open arcs.

On the other hand ARPES will find a gap along the pseudo Fermi surface, which, for $|u_{{\bf k}}|$ small compared to Fermi energy, will be very close to the Fermi surface of non-interacting fermions. ARPES cannot distinguish this gap from the superconducting gap; since there is no superconducting order in our ground state, it should be identified as a pseudogap. Indeed it suppresses the density of states and various thermodynamic response functions compared with non-interacting fermions or a Fermi liquid.

It should be emphasized that ARPES finds very broad electron spectral functions in the underdoped cuprates, with very small weight in a coherent quasiparticle peak (if present)\cite{Yoshida}. This means there is very strong scattering that render single particle excitations incoherent. This important piece of physics, which is related to the fact that underdoped cuprates are doped Mott insulators, is missing in our model.

Since our Fermi arcs are {\em not} closed, Luttinger's theorem (also known as Luttinger sum rule)\cite{AGD} is not satisfied the way in a Fermi liquid, where the Fermi surfaces are closed. This is, of course, just one of many indications that we have a non-Fermi liquid here. The way Luttinger sum rule gets satisfied here involves the combination of the Fermi arcs {\em and} the pseudo Fermi surfaces. This is because the fermion Green's function at zero energy, $G(\omega=0, {\bf k})$ changes sign across the pseudo Fermi surfaces (just like real Fermi surfaces), even though the fermion spectral weight there is {\em zero} (instead of a delta-function peak as for real Fermi surfaces of a Fermi liquid, and reflecting the pseudogap). Reversing the logic, Luttinger sum rule {\em requires} the co-existence of open Fermi arcs and pseudo Fermi surface(s) of the type discussed here.

\section{Quantum Oscillations}

The conclusions above follow straightforwardly from the exact solution of (\ref{eq:model}). In this section we discuss a more subtle question, namely how the system responds to a perpendicular magnetic field, which gives rise to quantum oscillations and represents another important probe of the pseudogap and Fermi arcs\cite{Proust,Sebastian}. Here we focus on the orbital coupling of the magnetic field, and ignore its (experimentally much weaker) Zeeman effect which, as discussed earlier, is straightforward to include.

We have three types of low-energy excitations in our model: (i) Spin-up fermions, whose dispersion is
\begin{eqnarray}
E^\uparrow_{{\bf k}}=\epsilon_{{\bf k}},
\end{eqnarray}
living near the black Fermi arcs of Fig. 1(b); (ii) Spin-down fermions, whose dispersion is
\begin{eqnarray}
E^\downarrow_{{\bf k}}=\epsilon_{{\bf k}} + u_{{\bf k}},
\end{eqnarray}
living near the red Fermi arcs of Fig. 1(b); and (iii) Singlet fermion pairs, whose dispersion is
\begin{eqnarray}
E^{\rm pair}_{2{\bf k}}=2\epsilon_{{\bf k}} + u_{{\bf k}},
\end{eqnarray}
living near the pseudo Fermi surfaces (blue) of Fig. 1(b). In the presence of a magnetic field ${\bf B}$, each of them propagate around their respective iso-energy contour following the semi-classical equations of motion\cite{Book}:
\begin{eqnarray}
\frac{d{\bf k}}{dt} &=& -\frac{q}{\hbar c}{\bf v}_{{\bf k}}\times {\bf B},\\
\hbar {\bf v}_{{\bf k}} &=& \nabla_{{\bf k}}(E_{{\bf k}}),
\label{eq:oscillation}
\end{eqnarray}
where $q=e$ for single fermions and $q=2e$ for the singlet pair.
If there were no interaction/restriction among them, these would be the contours illustrated in Fig. 1(a). Quantizing such periodic motion gives rise to Landau levels and density of states oscillating periodically with $1/B$ with a periodicity determined by the area of these contours\cite{Book}. The situation here is quite different: As each of them arrives at the point where $u_{{\bf k}}=0$, continuing with the trajectory given by their dispersion would take them into a region forbidden either by energy conservation or Pauli blocking. This is quite similar to a quantum mechanical particle following a semi-classical trajectory hitting an energy barrier. The resolution is, of course, that the particle quickly tunnels through the classically forbidden region and lands at the closest classically allowed point (due to the chirality imposed by the magnetic field back-scattering is also forbidden), and then continues with the semi-classical motion. This results in the momentum space trajectories illustrated in Fig. 2. Since the tunneling process is expected to be very quick, the periods of such motions are much shorter because the particles traverse only a small part of their semi-classical trajectories. As a result the quantum oscillations resulting from such trajectories have much smaller effective ``areas". If interpreted in the same way as quantum oscillations in Fermi liquids, one would conclude there exist Fermi ``pockets" that are much smaller than the original Fermi surface satisfying the Luttinger sum rule. This is indeed what is seen in the pseudogap region of the cuprate phase diagram\cite{Sebastian}.

\begin{figure}%[hbt]
%\vspace{2in}
\center
\includegraphics[width=0.4\textwidth]{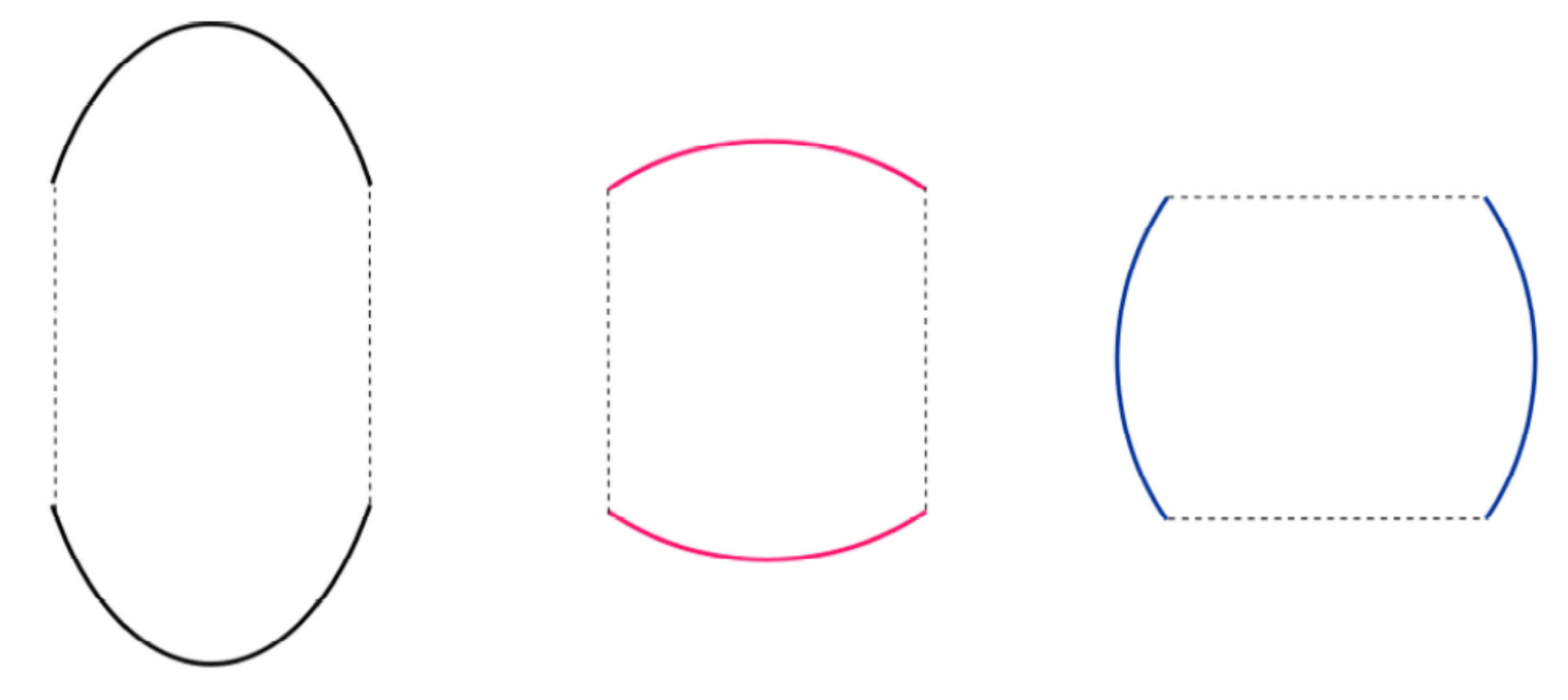}
\caption{The momentum space trajectories of the three types of low-energy excitations of our model. Solid lines represent semi-classical trajectories while dotted lines represent tunneling through forbidden region.}
\label{fig:fig2}
\end{figure}

We note among the many competing theories of small Fermi pockets\cite{Proust}, ours belongs to the group in which there is {\em no} broken translation symmetry (and the corresponding Fermi surface folding). Within this group the theory of Ref. \cite{Pereg-Barnea} is perhaps the closest to ours in spirit. We would like to emphasize Ref. \cite{Pereg-Barnea} models the pseudogap by a superconducting gap (which is argued to lose phase coherence at long length scales so that the system is {\em not} a superconductor); in our model the pseudogap has nothing to do with superconducting gap, and (as noted earlier) can be viewed as due to pairing fluctuation of the PDW type instead.

\section{Robustness of results and possible superconducting instability} 

In this section we demonstrate the robustness of the results that follow from the exact solution of (\ref{eq:model}) against more generic repulsive interactions. To this end we consider a perturbing Hubbard interaction:
\begin{eqnarray}
H' = J \sum_{i}n_{i\uparrow}n_{i\downarrow} = \frac{J}{N}\sum_{{\bf k}{\bf k}'{\bf q}}c^\dagger_{{\bf k}+{\bf q}\uparrow}c^\dagger_{{\bf k}'-{\bf q}\downarrow}c_{{\bf k}'\downarrow}c_{{\bf k}\uparrow},
\label{eq:Hubbard}
\end{eqnarray}
where $i$ is a site index, $J$ is Hubbard interaction strength (the unusual notation is going to be justified soon) and $N$ is system size. To linear order in $J$, we only need to evaluate $H'$ in the Fock subspace in which $H$ of (\ref{eq:model}) is degenerate:
\begin{eqnarray}
H' = (J/N)\sum_{({\bf k}{\bf k}')}[f(n_{\bf k}, n_{{\bf k}'}) - 2 {\bf S}_{\bf k}\cdot {\bf S}_{{\bf k}'}],
\label{eq:1st order Hubbard}
\end{eqnarray}
where ${\bf S}_{\bf k}$ is the spin operator associated with momentum ${\bf k}$, taking values $s=1/2$ for $n_{\bf k} =1$ and $s=0$ for $n_{\bf k} =0, 2$; $f(0, 0)=f(0, 1)=f(1,0)=0$, $f(1,1)=1/2$, $f(1,2)=f(2,1)=1$, and $f(2,2)=2$. The summation is over all pairs $({\bf k}{\bf k}')$. The ferromagnetic coupling in (\ref{eq:1st order Hubbard}) for $J > 0$ results from the {\em direct} exchange in momentum space; this is opposite to what happens in the usual Hubbard model at half-filling, where a super exchange in real space results in an antiferromagnetic coupling\cite{Book}. As anticipated earlier this ferromagnetic coupling fully polarizes the fermion spins in the $n_{\bf k} = 1$ region and removes the massive ground state degeneracy of (\ref{eq:model}). Due to the long-range nature of the spin-spin interaction (in momentum space) in (\ref{eq:1st order Hubbard}), there are {\em no} gapless spin-wave excitations here. The origin of such unusual behavior can be traced to the long-range nature of the $u_{\bf k}$ interaction (in {\em real} space) in (\ref{eq:model}). As a result this spontaneous magnetization has essentially the same effect as a Zeeman splitting. We thus expect our earlier results are robust for $J > 0$.

The situation is quite different for $J < 0$ (attractive Hubbard interaction). In this case the spin-spin interaction in (\ref{eq:1st order Hubbard}) is antiferromagnetic, and highly frustrated due to its long-range nature. Due to this frustration any state with ${\bf S}_{\rm tot} = \sum_{\bf k}{\bf S}_{\bf k} = 0$ is a ground state, and the massive ground state degeneracy is reduced but {\em not} lifted at linear order in $J$, and remains exponentially large in system size. The fate of the ground state is thus determined by higher-order processes in $J$. To this end we expect the BCS pairing interaction considered in Ref. \cite{Phillips} to dominate, resulting in a superconducting ground state, in whose spectrum the superconducting gap coexists with the pseudogap discussed earlier. We leave a detailed analysis of this case to future work.

\section{Summary and Discussions}

To summarize, we have presented a simple model that gives rise to Fermi arcs and pseudogap, in ways that are qualitatively consistent with phenomenology of underdoped cuprates, especially those of ARPES and quantum oscillation experiments. As emphasized earlier it is not our intention to present a comprehensive theory of cuprate phenomenology here. On the other hand we believe such a simple model is interesting in its own right, and may well contain the basic ingredients of the strong correlation physics responsible for the relevant phenomenology of cuprates and other systems.

We could attempt a more quantitative comparison with cuprate phenomenology, by choosing appropriate momentum, doping and perhaps even temperature dependence of parameters of our model. For example Fig. 1 corresponds to $\epsilon_{\bf k}$ and $u_{\bf k}$ with a $C_2$ symmetry, instead of $C_4$ symmetry that is appropriate to (some of the) cuprates. It is straightforward to enlarge their symmetry, which will result in four instead of two pairs of Fermi arcs. One can in principle force very detailed agreement, by fine-tuning $\epsilon_{\bf k}$ and $u_{\bf k}$ in ad hoc ways. On the other hand as stated before the main point of the present work is providing a proof of principle, and possible starting point for future work. Fine tuning would not add any additional insight, but might dilute our main message instead. On that note let us reiterate that there is also important cuprate physics missing in our model, especially those associated with the Mott insulator. In addition to those already mentioned we add that antiferromagnetism plays a very important role in cuprates, while in our model it appears to be suppressed; instead ferromagnetism is found with generic repulsive interactions. Understanding the possible interplay between Fermi arcs/pseudogap and Mott physics in the context of our model is a very interesting direction for future work.

\section*{Acknowledgments}
The author thanks Chandan Setty for correspondences that motivated the present work, and Joseph Ghobrial for graphics assistance.
This work was supported by the National Science Foundation Grant No. DMR-1932796, and performed at the National High Magnetic Field Laboratory, which is supported by National Science Foundation Cooperative Agreement No. DMR-1644779, and the State of Florida.


\begin{references}

\bibitem{Proust} For a recent review, see, {\em e.g.}, Cyril Proust and Louis Taillefer, The Remarkable Underlying Ground States of Cuprate Superconductors, Annual Review of Condensed Matter Physics
Vol. 10:409-429 (2019).

\bibitem{Yoshida} See, {\em e.g.}, Teppei Yoshida, Makoto Hashimoto, Inna M. Vishik, Zhi-Xun Shen, Atsushi Fujimori,
Pseudogap, Superconducting Gap, and Fermi Arc in High-Tc Cuprates Revealed by Angle-Resolved Photoemission Spectroscopy,
J. Phys. Soc. Jpn. {\bf 81}, 011006 (2012).

\bibitem{Book} See, {\em e.g.}, Steven M. Girvin and Kun Yang, {\it Modern Condensed Matter Physics}, ISBN: 9781107137394, Cambridge University Press, Cambridge (March 2019).

\bibitem{hk} Y. Hatsugai and M. Kohmoto, Exactly Solvable Model of Correlated Lattice Electrons in Any Dimensions, Journal of the Physical Society of Japan {\bf 61}, 2056 (1992).

\bibitem{Baskaran} G. Baskaran, An Exactly Solvable Fermion Model: Spinons, Holons and a non-Fermi Liquid Phase, Modern Physics Letters B {\bf 5}, 643 (1991).

\bibitem{Phillips}
P. W. Phillips, L Yeo, E. W. Huang, Exact theory for superconductivity in a doped Mott insulator, arXiv:1912.01008; Nat. Phys. (2020) https://doi.org/10.1038/s41567-020-0988-4.

\bibitem{Setty} Chandan Setty, Pairing instability on a Luttinger surface: A non-Fermi liquid to superconductor transition and its Sachdev-Ye-Kitaev dual,
Phys. Rev. B {\bf 101}, 184506 (2020).
%

\bibitem{Agterberg} Daniel F. Agterberg, J. C. Seamus Davis, Stephen D. Edkins, Eduardo Fradkin, Dale J. Van Harlingen, Steven A. Kivelson, Patrick A. Lee, Leo Radzihovsky, John M. Tranquada, Yuxuan Wang, The Physics of Pair Density Waves,
 Annual Review of Condensed Matter Physics {\bf 11}, 231 (2020).

\bibitem{Lee} Patrick A. Lee, Amperean pairing and the pseudogap phase of cuprate superconductors, Phys. Rev. X {\bf 4}, 031017 (2014); Zhehao Dai, T. Senthil, Patrick A. Lee, Modeling the pseudogap metallic state in cuprates: quantum disordered pair density wave, Phys. Rev. B {\bf 101}, 064502 (2020).

\bibitem{Shibauchi}
T. Shibauchi, L. Krusin-Elbaum, Ming Li, M. P. Maley, and P. H. Kes, Closing the Pseudogap by Zeeman Splitting in Bi2Sr2CaCu2O8+y at High Magnetic
Fields,
Phys. Rev. Lett. {\bf 86}, 5763 (2001).

\bibitem{AGD} See, {\em e.g.}, A. A. Abrikosov, L. P. Gorkov, and I. E. Dzyloshinski, Methods of Quantum Field Theory in Statistical Physics (Dover Books on Physics, 1963).

\bibitem{Sebastian} Suchitra E Sebastian, Neil Harrison and Gilbert G Lonzarich,
Towards resolution of the Fermi surface in
underdoped high-Tc superconductors, Rep. Prog. Phys. {\bf 75}, 102501 (2012).

\bibitem{Pereg-Barnea}
T. Pereg-Barnea, H. Weber, G. Refael, M. Franz, Quantum oscillations from Fermi arcs, Nature Physics {\bf 6}, 44 (2009).
\end{references}
\end{document}